\documentclass[%
 reprint,
superscriptaddress,
 amsmath,amssymb,
 aps,
pra, 
longbibliography,
floatfix ]{revtex4-1}

\usepackage{natbib} 
\usepackage{amsmath} 
\usepackage{graphicx} 
\usepackage[usenames,dvipsnames]{color}  
\usepackage{longtable}
\usepackage[caption=false]{subfig}
\usepackage{multirow}
\usepackage{amsmath}
\usepackage{epstopdf}
\usepackage[flushleft]{threeparttable}
\usepackage[english]{babel}
\usepackage{makecell}
\usepackage[section]{placeins}

\makeatletter
\def\bbl@set@language#1{%
  \edef\languagename{%
    \ifnum\escapechar=\expandafter`\string#1\@empty
    \else\string#1\@empty\fi}%
  \@ifundefined{babel@language@alias@\languagename}{}{%
    \edef\languagename{\@nameuse{babel@language@alias@\languagename}}%
  }%
  \select@language{\languagename}%
  \expandafter\ifx\csname date\languagename\endcsname\relax\else
    \if@filesw
      \protected@write\@auxout{}{\string\select@language{\languagename}}%
      \bbl@for\bbl@tempa\BabelContentsFiles{%
        \addtocontents{\bbl@tempa}{\xstring\select@language{\languagename}}}%
      \bbl@usehooks{write}{}%
    \fi
  \fi}
\newcommand{\DeclareLanguageAlias}[2]{%
  \global\@namedef{babel@language@alias@#1}{#2}%
} \makeatother

\DeclareLanguageAlias{en}{english}

\begin{document}

\title{Extending  Machine Learning to Predict Unbalanced Physics Course Outcomes}

\author{Seth DeVore}%
\author{Jie Yang}%
\author{John Stewart}
\email{jcstewart1@mail.wvu.edu}
\affiliation{%
Department of Physics and Astronomy, West Virginia University,
Morgantown WV, 26506
}%

\date{\today}

\begin{abstract}
Machine learning algorithms have recently been used to classify
students as those likely to receive an A or B or students likely
to receive a C, D, or F in a physics class. The performance
metrics used in that study become unreliable when the outcome
variable is substantially unbalanced. This study seeks to further
explored the classification of students who will receive a C, D,
and F and extend those methods to predicting whether a student
will receive a D or F. The sample used for this work ($N=7184$) is
substantially unbalanced with only 12\% of the students receiving
a D or F. Applying the same methods as the previous study produced
a classifier that was very inaccurate, classifying only 20\% of
the D or F cases correctly. This study will focus on the random
forest machine learning algorithm. By adjusting the random forest
decision threshold, the correct classification rate of the D or F
outcome rose to 46\%. This study also investigated the previous
finding that demographic variables such as gender,
underrepresented minority status, and first generation status had
low variable importance for predicting class outcomes.
Downsampling revealed that this was not the result of the
underrepresentation of these students. An optimized classification
model was constructed which predicted the D and F outcome with
46\% accuracy and C, D, and F outcome with 69\% accuracy; the
accuracy of prediction of these outcomes is called ``sensitivity''
in the machine learning literature. Substantial variation was
detected when this classification model was applied to predict the
C, D, or F outcome for underrepresented demographic groups with
61\% sensitivity for women, 67\% for underrepresented minority
students, and 78\% for first-generation students. Similar
variation was observed for the D and F outcome.

\end{abstract}

\maketitle

\section{Introduction}

Physics courses, along with other core science and mathematics
courses, form key hurdles for Science, Technology, Engineering,
and Mathematics (STEM) students early in their college career.
Student success in these classes is important to improving STEM
retention; the success of students traditionally underrepresented
in STEM disciplines in the core classes may be a factor limiting
efforts to increase inclusion in STEM fields. Physics Education
Research (PER) has developed a wide range of research-based
instructional materials and practices to help students learn
physics \cite{meltzer2012resource}. Research-based instructional
strategies have been demonstrated to increase student success and
retention \cite{freeman2014active}. While some of these strategies
are easily implemented for large classes, others have substantial
implementation costs. Further, no class could implement all
possible research-based strategies, and some may be more
appropriate for some subsets of students than for others. One
method to allow the direction of resources to students who would
most benefit would be to identify at-risk students early in
physics classes. The effective classification of students at risk
in physics classes represents a promising new research strand in
PER.

The need for STEM graduates continues to increase at a rate that
is outstripping STEM graduation rates across American
institutions. A 2012 report from the President's Council of
Advisors on Science and Technology \cite{olson2012engage}
identified the need to increase graduation of STEM majors to avoid
a projected shortfall of one million STEM job candidates over the
next decade. The U.S. Department of Education reported that STEM
attrition rates range from  59\% for computer/information science
majors to 38\% for math majors with an average of 48\%
\cite{chen2013}. With demand for jobs requiring at least a STEM
bachelors degree growing to 20\% of the job market over the last
decade \cite{nsb2015}, but with STEM degree completion rates
remaining only 40\% among students initially majoring in STEM
\cite{olson2012engage}, improvement in retention could relieve
some of the growing shortfall. Targeting interventions to students
at risk in core introductory science and mathematics courses taken
early in college offer one potential mechanism to improve STEM
graduation rates.

Improving STEM retention has long been an important area of
investigation for science education researchers \cite{rask2010,
chen2013, shaw2010, maltese2011,zhang2004,french2005,
marra2012,hall2015}. Many studies have shown that measures of
pre-college preparation (i.e. high school GPA and ACT or SAT
scores) in concert with college performance measures such as
college GPA are the strongly predictive of student success. With
introductory courses in physics, mathematics, and chemistry being
high attrition points for STEM majors, work focused on identifying
factors related to student success in these courses is key to
understanding the retention problem. In recent years, educational
data mining has become a prominent method of analyzing student
data to inform course redesign \cite{baepler2010, baker2009,
papamitsiou2014,dutt2017,romero2010}.

\subsection{Prior Study : Study 1}
This study extends and explores the results of Zabriskie {\it et
al.} \cite{zabriskie2019gpp} which will be referenced as Study 1
in this work. Study 1 used institutional data such as ACT scores
and college GPA (CGPA) as well as data collected within a physics
class such as homework grades and test scores to predict whether a
student would receive an A or B in the first and second semester
calculus-based physics classes at a large university. The study
used both logistic regression and random forests to classify
students. Random forest classification using only institutional
variables was 73\% accurate for the first semester class. This
accuracy increased to 80\% by the fifth week of the class when
in-class variables were included. The logistic regression and
random forest classification algorithms generated very similar
results. Study 1 chose to predict A and B outcomes, rather than
the more important A, B, and C outcomes, partially because the
sample was significantly unbalanced. Sample imbalance makes
classification accuracy more difficult to interpret. The study
also made a number of decisions about classification parameters
such as the relative size of the test and training dataset, the
number of trees grown, and the decision threshold (explained in
Sect. \ref{sec:methods}) which should be further explored. Study 1
also investigated the effect of a number of demographic variables
on grade prediction (gender, underrepresented minority status, and
first generation status) and found they were not important to
grade classification. These groups were very underrepresented in
the courses studied; it was unclear as to what extent sample
imbalance caused by underrepresentation was the cause of the low
importance of the demographic variables.

\subsection{Research Questions}
This study seeks to more fully explore the random forest machine
learning algorithm and explore questions left unanswered by Study
1. It also seeks to extend the application of the algorithm to
unbalanced dependent and independent variables.

    This study seeks to answer the following research questions:
\begin{enumerate}
    \item[RQ1:] How can machine learning algorithms be applied to predict unbalanced physics class outcomes?
    \item[RQ2:] What is a productive set of performance metrics to characterize the classification algorithms?
    \item[RQ3:] What sample size is required for accurate prediction of physics class outcomes?
    \item[RQ4:] How does classification accuracy differ for groups underrepresented in physics? How can machine learning
    models be optimized to predict the outcomes of all groups with equal accuracy?
\end{enumerate}

\subsection{Educational Data Mining}
Educational Data Mining (EDM) can be described as the use of
statistical, machine learning, and traditional data mining methods
to draw conclusions from large educational datasets while
incorporating predictive modeling and psychometric modeling
\cite{romero2010}. In a 2014 meta-analysis of 240 EDM articles by
Pe{\~n}a-Ayala, 88\% were found to use a statistical and/or
machine learning approach to draw conclusions from the data
presented. Of these studies 22\% analyzed student behavior, 21\%
examined student performance, and 20\% examined assessments
\cite{pena2014educational}. Pe{\~n}a-Ayala also found that
classification was the most common method used in EDM applied in
42\% of all analyses, with clustering used in 27\%, and regression
used in 15\% of studies.

Educational Data Mining encompasses a large number of statistical
and machine learning techniques with logistic regression, decision
trees, random forests, neural networks, naive Bayes, support
vector machines, and K-nearest neighbor algorithms commonly
applied \cite{romero2008data}. Pe{\~n}a-Ayala's
\cite{pena2014educational} analysis found 20\% of studies employed
Bayes theorem and 18\% decision trees. Decision trees and random
forests are one of the more commonly used techniques in EDM making
them suitable techniques to investigate our research questions and
explore techniques to assess the success of machine learning
algorithms. More information on the fundamentals of these and
other machine learning techniques are readily available through a
number of machine learning texts \cite{james2017,
muller2016introduction}

\subsection{Grade Prediction and Persistence}

\renewcommand{\tabcolsep}{2mm}
\begin{table*}[!htb]
    \caption{\label{tab:variables} Full list of variables.}
    \begin{center}

        \begin{tabular}{lcl}

            Variable  &  Type & Description                                                                                  \\\hline
            Gender    &Dichotomous&  Gender (Men = 1 Women = 0).                                                        \\
            URM       &Dichotomous&  Student does not identify as White non-Hispanic or Asian (True = 1, False = 0). \\
            CalReady &Dichotomous&   The first math class taken calculus or more advanced (True = 1, False = 0).  \\
            FirstGen  &Dichotomous&  Student is a first generation college student (True = 1, False = 0).                                 \\
            CmpPct    &Continuous&  Percentage of hours attempted that were completed at the start of course. \\
            CGPA      &Continuous&  College GPA at start of course.                                                       \\
            STEMHrs   &Continuous&  Number of STEM (Math, Bio, Chem, Eng, Phys) credit hours completed at start of course.       \\
            HrsCmp    &Continuous&  Total credits hours earned at start of course.                                                                \\
            HrsEnroll &Continuous&  Total credits hours enrolled at start of course.                                                           \\
            HSGPA     &Continuous&  High school GPA.                                                                              \\
            ACTM      &Continuous&  ACT/SAT mathematics percentile.                                                                      \\
            ACTV      &Continuous&  ACT/SAT verbal percentile.                                                                    \\
            APCredit  &Continuous&  Number of credits hours received for AP tests.                                                     \\
            TransCrd  &Continuous&  Number of credits hours received for transfer courses.                                             \\ \\
        \end{tabular}
    \end{center}
\end{table*}
\renewcommand{\tabcolsep}{2pt}

While EDM is used for a wide array of purposes, it has often been
used to examine student performance and persistence. One survey by
Shahiri {\it et al.} summarized 30 studies in which student
performance was examined using EDM techniques
\cite{shahiri2015review}. Neural networks and decision trees were
the two most common techniques used in studies examining student
performance with naive Bayes, K-nearest neighbors, and support
vector machines used in some studies. A study by Huang and Fang
examined student performance on the final exam for a
large-enrollment engineering course using measurements of college
GPA, performance in 3 prerequisite math classes as well as Physics
1, and student performance on in-semester examinations
\cite{huang2013predicting}. They analyzed the data using a large
number of techniques commonly used in EDM and found relatively
little difference in the accuracy of the resulting models. Study 1
also found little difference in the performance of machine
learning algorithms in predicting physics grades. Another study
examining an introductory engineering course by Marbouti {\it et
al.} used an array of EDM techniques to predict student grade
outcomes of C or better \cite{marbouti2016models}. They used
in-class measures of student performance including homework, quiz,
and exam 1 scores and found that logistic regression provided the
highest accuracy at 94\%. A study by Macfadyen  and Dawson
attempted to identify students at risk of failure in an
introductory biology course \cite{macfadyen2010mining}. Using
logistic regression they were able to identify students failing
(defined as having a grade of less than $50\%$) with 81\%
accuracy. Interest in grade prediction and persistence in STEM
classes has risen to the point where many universities are using
EDM techniques to improve retention of STEM students
\cite{bin2013overview}.

The use of machine learning techniques in physics classes has only
begun recently. Beyond Study 1, random forests were used in a 2018
study by Aiken {\it et al.} to predict student persistence as
physics majors and identify the factors that are predictive of
students either remaining physics majors or becoming engineering
majors \cite{aiken2018modeling}.

\section{Methods}
\label{sec:methods}

\subsection{Sample}
\label{sec:context}

This study was performed using course grades from the
introductory, calculus-based mechanics course (Physics 1) taken by physical
science and engineering students at a large eastern
land-grant university serving approximately 30,000 students. The
general university undergraduate population had ACT scores ranging
from 21 to 26 (25th to 75th percentile) \cite{usnews}. The overall
undergraduate demographics were 80\% White, 4\% Hispanic, 6\%
international, 4\% African American, 4\% students reporting with
two or more races, 2\% Asian, and other groups each with 1\% or
less \cite{usnews} The sample was primarily male (82\%).

The sample for this study was drawn from institutional records and
includes all students who completed Physics 1 from 2000 to 2018, a
total of 7184 students. Over the period studied, the instructional
environment of the course varied widely, and as such, the results
of this study may provide a general picture of the performance of
machine learning algorithms to predict physics grades. Improved
performance might arise within more stable instructional
environments. Prior to the spring 2011 semester, the course was
presented traditionally with multiple instructors teaching largely
traditional lectures and students performing cookie-cutter
laboratory exercises. In spring 2011, the department made a
commitment to improved instruction with the implementation of a
Learning Assistant (LA) program \cite{otero2010physics} and the
hire of an expert educator to manage the program. This educator
brought the Peer Instruction pedagogy to the lecture component of
the course \cite{crouch2001peer}. Learning Assistants were
instructed in reformed pedagogy and presented lessons from the
University of Washington {\it Tutorials in Introductory Physics}
\cite{mcdermott1998} in the laboratory sections; students also
continued to perform traditional laboratory experiments in the
same sessions.  In fall 2015, the program was modified because of
a change in funding with LAs assigned to only a subset of
laboratory sections. The course introduced a team-teaching model
at this time featuring strong coordination of the lecture and
laboratory components; there was little coordination between the
lecture and laboratory components prior to fall 2015. The
Tutorials were replaced with open source materials
\cite{opensource} which lowered textbook cost to students and
allowed full integration of the research-based materials with the
laboratory activity.

\subsection{Variables}

The variables used in this study were drawn from institutional
records and are shown in Table \ref{tab:variables}. Two types of
variables are used: two-level dichotomous variables and continuous
variables. A few variables require additional explanation. The
variable CalReady measures the student's math-readiness. Calculus
1 is a pre-requisite for Physics 1. For the vast majority of
students in Physics 1, the student's four-year degree plans assume
the student enrolls in Calculus 1 their first semester at the
university. These students are considered ``math ready.'' A
substantial percentage of the students at the institution studied
are not math ready. Study 1 used a 3-level math-readiness
variable; this study uses a 2-level variable to allow a more
thorough exploration of unbalanced dichotomous independent
variables. The variable STEMHrs captures the number of credit
hours of STEM classes completed before the start of the course
modeled. STEM classes include mathematics, biology, chemistry,
engineering, and physics classes.

Demographic information was also collected from institutional
records. Students self-report first-generation status; students
are considered first generation if neither of their parents
completed a four-year degree. Racial and ethnic information was
also accessed. A student was classified as an underrepresented
minority student (URM) if they did not reported ethnicity of
Hispanic or reported a race other than White or Asian. Gender was
also collected from university records; for the period studied
gender was recorded as a binary variable by the institution. While
not optimal, this reporting is consistent with the use of gender
in most studies in PER; for a more nuanced discussion of gender
and physics see Traxler {\it et al.}
\cite{traxler_enriching_2016}.

\subsection{Random Forest Classification Models}

This work employs the random forest machine learning algorithm to
predict students' final grade outcomes in introductory physics.
Random forest are one of many machine learning classification
algorithms. Study 1 reported that most machine learning algorithms
had similar performance when predicting physics grades. A
classification algorithm seeks to divide a dataset into multiple
classes. This study will classify students as those which will
receive an A or B (AB) and students who will receive a C, D, or F
(CDF) in Physics 1 following Study 1. It will also classify
students who will receive an A, B, or C (ABC) and students who
will receive a D or F (DF). This classification is fairly
unbalanced and will require additional techniques.

To understand the performance of a classification algorithm, the
dataset is first divided into test and training datasets. The
training dataset is used to develop the classification model, to
train the model. The test dataset is then used to characterize the
model. The classification model is used to predict the outcome of
each student in the test dataset; this prediction is compared to
the actual outcome. Section \ref{sec:pred} discusses performance
metrics used to characterize the success of the classification
algorithm.

The random forest algorithm uses decision trees, another machine
learning classification algorithm. Decision trees work by
splitting the dataset into two or more subgroups based on one of
the model variables. The variable selected for each split is
chosen to divide the dataset into the two most homogeneous subsets
of outcomes possible, that is, subsets with a high percentage of
one of the two classification outcomes. The variable and the
threshold for the variable represents the decision for each node
in the tree. For example, one node may split the dataset using the
criteria (the decision) that a student's college GPA is less than
3.2. The process continues by splitting the subsets forming the
decision tree until each node contains only one of the two
possible outcomes. Decision trees are less susceptible to
multicollinearity than many statistical methods common in PER such
as linear regression \cite{breiman1984classification}.

Random forests  extend the decision tree algorithm by growing many
trees instead of a single tree. The ``forest'' of decision trees
is used to classify each instance in the data; each tree ``votes''
on the most probable outcome. The decision threshold determines
what fraction of the trees must vote for the outcome for the
outcome to be selected as the overall prediction of the random
forest. Random forests use bootstrapping to prevent one variable
from being obscured by another variable. Individual trees are
grown on $Z$ subsamples generated by sampling the training data
set with replacement. Each of these samples is fit using a subset
of size $m$ of the variables, $m=\sqrt{k}$, where $k$ is the
number of independent variables in the model \cite{hastie2009}.
This method ensures the trees are not correlated and that the
stronger variables do not overwhelm weaker variables
\cite{james2017}. The ``randomForest'' package in ``R'' was used
for the analysis. This package provides a measure variable
importance, the mean decrease in accuracy \cite{liaw2002}. The
mean decrease in accuracy is the average decrease in
classification accuracy if the variable is removed
\cite{hastie2009}. This work uses bootstrapping to produce similar
variable importance measures for other performance metrics.

\subsection{Performance Metrics} \label{sec:pred}

The confusion matrix \cite{fawcett2006introduction} as shown in
Table \ref{confmat} summarizes the results of a classification algorithm
and is the basis for calculating most model performance metrics. To construct
the confusion matrix, the classification model developed from the training
dataset is used to classify students in the test dataset. The confusion matrix
categorizes the outcome of this classification.

\begin{table}[!htb]
    \caption{\label{confmat} Confusion Matrix}
    \begin{tabular}{c | c | c }
        &Actual Negative& Actual Positive \\\hline
        Predicted Negative& True Negative (TN) & False Negative (FN) \\\hline
        Predicted Positive& False Positive (FP)& True Positive
        (TP)\\
    \end{tabular}

\end{table}

For classification, one of the dichotomous outcomes is selected as
the positive result. In the current study, we use the DF or CDF
outcomes as the positive result. This choice was made because some
the the model performance metrics focus on the positive results
and we feel that most instructors would be more interested in
accurately identifying students at risk of failure.

From the confusion matrix, many performance metrics can be
calculated.  Study 1 reported the classification accuracy, the
fraction of correct predictions, shown in Eqn. \ref{eqn:acc}.
\begin{equation}
\label{eqn:acc} \mbox{Accuracy} = \frac{\mbox{TN} +
    \mbox{TP}}{N_{\mathrm{ test}}}
\end{equation}
where $N_{\mathrm{ test}}=\mbox{TP+TN+FP+FN}$ is the size of the
test dataset.

Sensitivity, the true positive rate (TPR), and specificity, the
true negative rate (TNR), characterize the rate of making accurate
predictions of either the positive or negative class. Sensitivity
is the fraction of the positive cases that are classified as
positive (Eqn \ref{eqn:sen}).
\begin{equation}
\label{eqn:sen} \mbox{Sensitiviy} = \mbox{TPR}
    = \frac{\mbox{TP}}{
        \mbox{TP}+
        \mbox{FN}
        }
    = \frac{\mbox{TP}}{N_{\mathrm{pos}}}
\end{equation}
where $N_{\mathrm{pos}}=\mbox{TP}+\mbox{FN}$ is the number of
positive cases in the test dataset. Specificity is the fraction of
the negative cases that are classified as negative (Eqn
\ref{eqn:spec}).
\begin{equation}
\label{eqn:spec} \mbox{Specificity} =\mbox{TNR}= \frac{\mbox{TN}}{
    \mbox{TN}+
    \mbox{FP}
    }
= \frac{\mbox{TN}}{N_{\mathrm{neg}}}
\end{equation}
where $N_{\mathrm{neg}}=\mbox{TN}+\mbox{FP}$ is the number of
negative cases in the test dataset.

Sensitivity and specificity can be adjusted by changing the
strictness of the classification criteria. If the model classifies
even the only slightly promising cases as positive, it will
probably classify most actually positive cases as positive
producing a high sensitivity. It will also make a lot of mistakes;
the precision or the the positive predictive value (PPV) captures
the rate of making correct predictions and is  defined as the
fraction of the positive predications which are correct (Eqn.
\ref{eqn:prec}).
\begin{equation}
\label{eqn:prec} \mbox{Precision} =\mbox{PPV}= \frac{\mbox{TP}}{
    \mbox{TP}+
    \mbox{FP}
    }
\end{equation}
This study will seek models that balance sensitivity and
precision; however, the correct balance for a given application
must be selected based on the individual features of the
situation. If there is little cost and no risk to an intervention,
then optimizing for higher sensitivity might be the correct choice
to identify as many students in the positive class as possible. If
the intervention is expensive or carries risk, optimizing the
precision so that most students who are given the intervention are
actually at risk might be more appropriate.

One challenge of applying machine learning methodologies to answer
academic questions in PER is that some terms that have well
established meanings in physics such as precision and accuracy
have been used differently in computer science. In what follows,
we will use the traditional meaning of precision as how well a
quantity is known, calling the computer science precision, PPV.
Accuracy will be as defined in Eqn. \ref{eqn:acc}.

Beyond simply evaluating the overall performance of a
classification algorithm, we would like to establish how much
better the algorithm performs than pure guessing. The sample used
in this study is substantially unbalanced between the DF or ABC
outcomes with 88\% of the students receiving an A, B, or C. If a
classification method guessed that all student would receive an A,
B, or C (the negative outcome), then the classifier would have a
sensitivity of 0, a specificity of 1, a PPV of 0, and an accuracy
of $0.88$. If the classifier guessed all students would receive an
D or F, the sensitivity would be 1, specificity 0, PPV 0.12, and
accuracy 0.12.

Additional performance metrics have been constructed to provide a
more complete picture of model performance. Cohen's kappa,
$\kappa$, measures agreement among observers \cite{cohen}
correcting for the effect of pure guessing as shown in Eqn.
\ref{eqn:kappa}.
\begin{equation}
\label{eqn:kappa} \kappa = \frac{p_0-p_e}{1-p_e}
\end{equation}
where $p_0$ is the observed agreement and $p_e$ is agreement by chance. Fit criteria
have been developed for $\kappa$ with
$\kappa$ less than $0.2$ as poor agreement, 0.2 to 0.4 fair
agreement, 0.4 to 0.6 moderate agreement, 0.6 to 0.8  good
agreement, and 0.8 to 1.0 excellent agreement between observers
\cite{altman1990practical}.

The Receive Operating Characteristic (ROC) curve (originally
developed to evaluate radar) plots the true positive rate (TPR) or
sensitivity against the false positive rate (FPR);
$\mbox{FPR}=1-\mbox{Specificity}$. The Area Under the Curve (AUC)
is a measure of the model's discrimination between the two
outcomes; AUC is the integrated area under the ROC curve. For a
classifier that uses pure guessing, the ROC curve is a straight
line between (0,0) and (1,1) and the AUC is $0.5$. An AUC of 1.0
represents perfect discrimination
\cite{hosmer2013applied,fawcett2006introduction}. Hosmer
\textit{et al.} \cite{hosmer2013applied}  suggest an AUC threshold
of $0.80$ for excellent discrimination. Study 1 provided examples
of ROC curves in the Supplemental Material.

Two other metrics attempt to balance multiple performance
measures. The $F_1$ metric is the harmonic mean of the precision
and sensitivity (the positive predictive value and the true
positive rate) shown in Eqn. \ref{eqn:F1value}.

\begin{equation}
\label{eqn:F1value} \frac{1}{F_1} =
\frac12\bigg(\frac{1}{\mbox{Sensitivity}}+\frac{1}{\mbox{PPV}}\bigg)
\end{equation}
As with the addition of parallel resistors, $F_1$ gives a stronger
weight to the smaller of the sensitivity and PPV.

The $g_{\mathrm{mean}}$ metric is the geometric mean of
sensitivity and specificity as shown in Eqn. \ref{eqn:gmean}.
\begin{equation}
\label{eqn:gmean} g_{\mathrm{mean}}=
\sqrt{\mbox{Sensitivity}\cdot\mbox{Specificity}}
\end{equation}

\subsection{Unbalanced Datasets}

This study, as well as Study 1, used a number of dichotomous
independent variables: Gender, FirstGen, URM, and CalReady. Each
variable further divides both outcome classes. The division of the
groups defined by these variables over the outcome variables is
shown in Table \ref{tab:demogroups}.

The outcomes in the dataset used in this study are unbalanced;
there are more students in the negative class (AB or ABC) than the
positive class (CDF or DF); this imbalance is severe for the DF
class. Imbalance in the training data  can cause learning
algorithms to perform poorly on the minority class
\cite{he2009learning,
chawla2004special,chawla2009data,sun2009classification,kotsiantis2006handling}.
To improve classification of the minority class, many different
forms of resampling have been introduced. Random undersampling or
downsampling balances the two classes by randomly eliminating
majority class examples. Downsampling, however, reduces the
overall training dataset size which may reduce overall
classification performance.

Random oversampling or upsampling also balances the two classes by
randomly duplicating minority class instances. This method is
susceptible to overfitting because duplicating records causes the
students who were duplicated to have more weight in the
classification process than other students. More sophisticated
upsampling methods have been constructed. Synthetic Minority
Oversampling Technique (SMOTE) \cite{chawla2002smote} generates
new integrated minority cases rather than copying existing cases.
It forms new minority case examples by interpolating existing
examples that are near each other in the parameter space. In
addition to creating a balanced dataset, cost-sensitive learning
methods also can be used improve performance with unbalanced
datasets
\cite{elkan2001foundations,ting2002instance,he2009learning}.

\subsection{Bootstrapping}
Bootstrapping is a computational method designed to eliminate
distributional assumptions (and their violation) common in
statistical methods. This study applies bootstrapping by creating
randomly selected subsets of the full sample. This allows the
uncertainty in performance metrics to be calculated. The random
forest algorithm internally applies bootstrapping growing a number
of trees selected by the user. For the evaluation of test and
training dataset size in Sec. \ref{sec:train}, 1000 bootstrap
replications were used with each growing 1000 decision trees for a
total of 1,000,000 decision trees per data point. This was
computationally very expensive. Examination of the standard errors
of this analysis and considering the small number of independent
variables suggested that a less conservative selection of
parameters would be appropriate. For the remainder of the
analysis, 200 bootstrap replications growing 200 decision trees
were used for a total of 40,000 trees per data point.

\subsection{Standard Deviation and Standard Error}
All tables and figures report the standard deviation of the
performance metrics calculated for the set of bootstrap
replications; error bars in figures are one standard deviation
long. These measure the variation between multiple subsamples of
the same dataset and provide a measure of the variation that
should be expected as the classification model is applied to new
data. In practice, the classification model would be constructed
from some sample of past students, then applied to predict the
outcomes of a new set of students. However, when comparing
differences in performance metrics or determining if variable
importance is different than zero, the standard error of the mean
should be used. The standard error divides the standard deviation
by the square root of the number of observations used to calculate
the standard deviation. For the test-train dataset size evaluation
in Sec. \ref{sec:train}, the standard error is the standard
deviation divided by $\sqrt{1000}=31.6$; for other calculations,
the standard error is the standard deviation divided by
$\sqrt{200}=14.1$.

\section{Results}
\renewcommand{\tabcolsep}{2mm}
\begin{table*}[!htb]
        \centering
        \begin{tabular}{| l| c c ccc| }\hline
            &N&Physics Grade&ACT Math \%&HSGPA & CGPA\\\hline
            Overall     &7184       &$2.70\pm1.3$   &$79\pm14$      &$3.71\pm0.5$       &$3.18\pm0.5$\\\hline
            AB Students &4506       &$3.48\pm0.5$   &$83\pm13$      &$3.84\pm0.4$       &$3.38\pm0.5$\\
            CDF Students&2678       &$1.38\pm1.0$   &$74\pm15$      &$3.50\pm0.5$       &$2.84\pm0.5$\\\hline
            ABC Students&6337       &$3.05\pm0.8$   &$80\pm14$      &$3.75\pm0.4$       &$3.25\pm0.5$\\
            DF Students &847        &$0.05\pm0.9$   &$73\pm15$      &$3.43\pm0.5$       &$2.65\pm0.5$\\\hline
            Women       &1270       &$2.83\pm1.2$   &$79 \pm 14$    & $ 3.94\pm 0.4$    & $3.38 \pm 0.5$\\
            Men         &5914       &$2.67 \pm 1.3$ &$79 \pm 14$    &$3.66 \pm 0.5$     &$ 3.14\pm 0.5$\\\hline
            URM         &388        &$2.42 \pm 1.3$ &$73 \pm 17$  &$3.53 \pm 0.5$     &$ 3.03\pm 0.6$\\
            Not URM     &6796       &$2.71 \pm 1.3$ &$80 \pm 14$    &$3.72 \pm 0.5$     &$ 3.19\pm 0.5$\\\hline
            First Gen.  &815        &$2.66\pm1.3$   &$77\pm15$      &$3.72\pm0.5$       &$3.15\pm0.5$\\
            Not First Gen.&6369     &$2.70\pm1.3$   &$80\pm14$      &$3.71\pm0.5$       &$3.18\pm0.5$\\\hline
            Calc. Ready &5622       &$2.84\pm1.2$   &$83\pm11$      &$3.78\pm0.4$       &$3.23\pm0.5$\\
            Not Calc. Ready&1562    &$2.20\pm1.2$   &$65\pm14$      &$3.47\pm0.4$       &$3.01\pm0.5$\\\hline
        \end{tabular}
    \caption{\label{tab:demo} Descriptive Statistics. All values are the mean $\pm$ the standard deviation. }
\end{table*}
\renewcommand{\tabcolsep}{2pt}

\renewcommand{\tabcolsep}{2mm}
\begin{table*}[!htb]
        \centering
        \begin{tabular}{| l| cc|cc|cc|cc| }\hline

\multicolumn{9}{|c|}{Predicting CDF}\\\hline
&\multicolumn{2}{|c}{CalReady}&\multicolumn{2}{|c}{FirstGen}&\multicolumn{2}{|c}{Gender}&\multicolumn{2}{|c|}{URM}\\\hline
Outcome &Cal Ready      &Not Cal Ready          &First Gen
&Not First Gen          &Men        &Women          &URM    &Not
URM\\\hline
AB      &3828           &678                    &494                &4012                   &3650       &856            &207    &4299   \\
CDF     &1794           &884                    &321
&2357                   &2264       &414            &181    &2497
\\\hline \multicolumn{9}{|c|}{Predicting DF}\\\hline
&\multicolumn{2}{|c}{CalReady}&\multicolumn{2}{|c}{FirstGen}&\multicolumn{2}{|c}{Gender}&\multicolumn{2}{|c|}{URM}\\\hline
ABC     &5065           &1272                   &725                 &5612                   &5187       &1150           &326    &6011\\
DF      &557            &290                    &90
&757                   &727         &120           &62
&785\\\hline
        \end{tabular}
    \caption{\label{tab:demogroups} Demographic composition of sample. Each entry shows the number of students in each subgroup.  }
\end{table*}
\renewcommand{\tabcolsep}{2pt}
The purpose of this work is to further understand the
classification of students who will receive a CDF outcome in
Physics 1 as was done in Study 1 and to extend this work to the
prediction of the more unbalanced DF outcome. Either dichotomous
outcome variable divides the sample into two subsets with
different academic characteristics. Table \ref{tab:demo} presents
overall academic performance measures for each outcome; the
variables are defined in Table \ref{tab:variables}. The
dichotomous independent variables further divide the subsets
defined by the outcome variables. The overall demographic
composition of the sample is shown in Table \ref{tab:demogroups}.

\subsection{Training and Test Dataset Size Requirements}
\label{sec:train} For any quantitative analysis, it is important
to ensure that a sufficient sample size is available to draw
accurate conclusions. In machine learning, a large training
dataset provides the learning algorithm with more unique cases
from which to learn; model performance generally increases with
training sample size \cite{scikit}. As with most analyses, the
precision with which the results are known increases with sample
size.  To determine how the precision of the model performance
metrics change with training dataset size, the original training
dataset was randomly downsampled to produce smaller training
datasets while the test dataset was held fixed. The sample was
first split into a test and training dataset where each
represented 50\% of the original sample. This provides a large
sample to train the algorithm and a large dataset to precisely
characterize the model produced.

\begin{figure*}[!htb]
    \centering
    \includegraphics[width=7in]{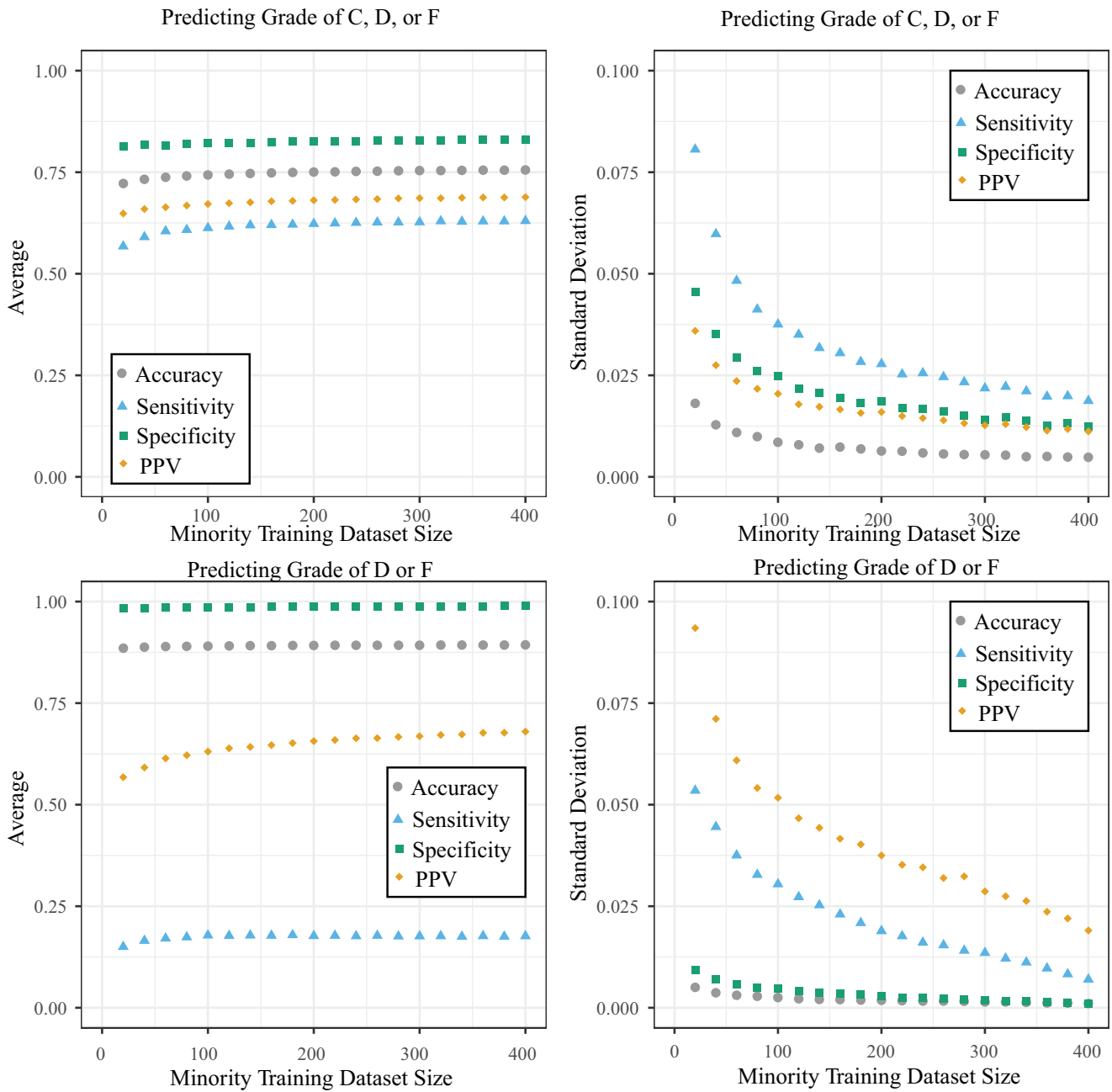}
    \caption{\label{fig:train} Model performance parameters as a function of the size of the minority outcome (CDF or DF) in the training dataset.  }
\end{figure*}

Figure \ref{fig:train} plots the minority outcome size in the
training dataset against the model accuracy, sensitivity,
specificity, and PPV as well as the standard deviations of these
quantities. We expect the smaller subdivision of the outcome
variable (the minority outcome) to be most important in
determining precision, and therefore, precision is examined in
terms of the minority outcome sample size. As expected, the
standard deviation decreases as sample size increases. For all
performance metrics, there is a weak increase  up to a minority
outcome sample size of approximately 100 with all performance
measures becoming approximately constant above his value.

Study 1 commented on a higher than desirable false negative rate
in its Limitations section (Study 1 coded the CDF outcome as
negative). One can clearly see this effect in Fig. \ref{fig:train}
where the sensitivity predicting the CDF outcome is approximately
60\% while the specificity is 80\%. The model predicts the CDF
outcome substantially less effectively than the AB outcome. This
effect becomes severe for the DF outcome with a sensitivity of
approximately 20\%. For both the DF and CDF outcomes, the PPV is
approximately 65\%; therefore, 65\% of the students classified as
earning a DF or CDF actually do.

The standard deviation curves in Fig. \ref{fig:train} are somewhat
different; this may have resulted from a ceiling or floor effect
for some performance metrics limiting the standard deviations. The
sample sizes required to achieve a desired precision are
commensurate; for the CDF outcome, to achieve a precision of 0.025
in the sensitivity, 220 students are required in the minority
outcome; for the DF outcome, 140 students are required for the
same precision.

A similar analysis was performed for the test dataset; the model
performance plots are shown in the Supplemental Material
\cite{supp}. The relation of the performance metrics to the test
dataset size was somewhat different than those of the training
dataset. The test dataset size had no effect on the average value
of the performance metric. This was to be expected; increasing the
test dataset size does not provide the learning algorithm with
additional unique cases to improve the prediction algorithm.
Slightly larger test datasets were required to achieve the same
level of precision as the training dataset shown in Fig.
\ref{fig:train}. For predicting the CDF outcome, 260 students were
required in the minority outcome test dataset to produce a
sensitivity with standard deviation 0.025; for the DF outcome, 160
students were required. The uncertainty of PPV was much larger
than the other quantities for the DF outcome.

Originally, we sought to provide some guidelines on minimum
required sample size and optimal test-train split ratio. We
abandoned this goal because bootstrapping can provide confidence
intervals and standard deviations for any quantity desired. The
required sample size then reverts to the traditional decision in
research, how much precision is required for the conclusions one
wishes to draw from the analysis. We will find that the test-train
split is controlled by the need to retain a maximum number of the
minority class of the dichotomous independent variables as
discussed in Sec. \ref{sec:optimal}.

The plots in Fig. \ref{fig:train} also suggest that rather than
focussing on a single overall performance parameter such as AUC or
$\kappa$ as was done in Study 1, that, for grade prediction, it
may be more productive to focus on optimizing multiple measures
simultaneously. We report sensitivity, specificity, accuracy, and
PPV as the models are optimized and report $\kappa$, AUC,
$g_{\mathrm{mean}}$, and $F_1$ only for the optimized models.

\subsection{Unbalanced Dependent Variables}

Both outcome variables, predicting CDF or DF, are unbalanced as
shown in Table \ref{tab:demo}; there are more students in one of
the classifications than the other. The CDF outcome is somewhat
unbalanced with 37\% of the students receiving a C, D, or F. The
DF outcome is quite unbalanced with 12\% of the students receiving
a D or F. Sample imbalance can produce a classifier that predicts
the outcomes of the majority and minority class with differing
precision. Multiple methods exist to correct for sample imbalance:
downsampling, upsampling, and hyperparameter tuning.

\subsubsection{Downsampling}
\begin{figure}[!h]
    \centering
    \includegraphics[width=3.5in]{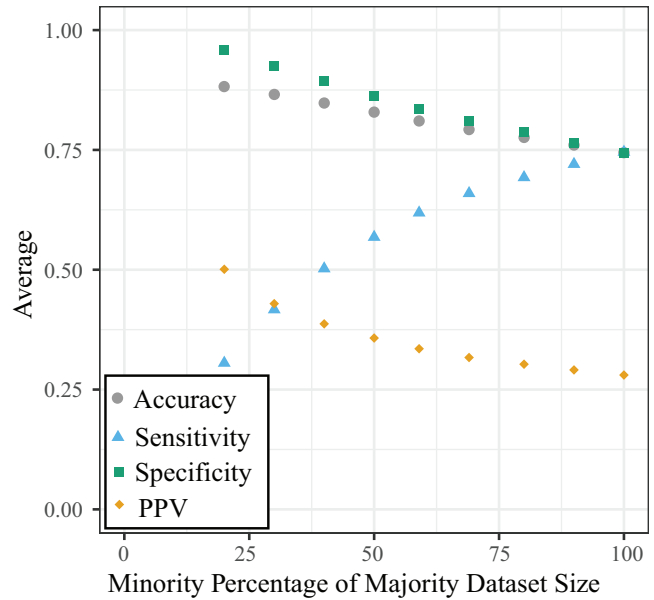}
    \caption{\label{fig:downsample}
    Model performance parameters as the majority training dataset is downsampled to
    minority dataset size predicting the DF outcome. The standard deviation did not exceed $0.02$ for any data point.  }
\end{figure}

Figure \ref{fig:train} shows that random forest models are less
effective at predicting CDF and DF outcomes; the  sensitivity
measures the fraction of CDF or DF outcomes that are correctly
predicted. One possible cause of this is the sample imbalance
which provides the random forest learning algorithm more examples
of the majority class, thus optimizing the model to correctly
identify these cases. One possible method to improve the
prediction of the minority results is downsampling; reducing the
size of the majority dataset by randomly sampling it without
replacement. Because downsampling reduces overall sample size, the
reduced sample still needs to meet the sample size requirements
explored in the previous section. Figure \ref{fig:downsample}
shows the effect of downsampling on the model performance metrics
predicting the DF outcome. The horizontal axis plots the
percentage ratio of the majority sample to the minority sample;
the two samples are balanced when the ratio is 100\%. The figure
clearly shows that as the majority class is downsampled overall
model accuracy and the correct prediction of the majority outcome
(specificity) decreases; however, the rate of correctly predicting
of the minority outcome dramatically increases. The PPV, the
fraction of correct positive predictions, also decreases with
downsampling. This may be a result of less data being provided to
train the algorithm resulting in more incorrect classifications.

If a balance of sensitivity and specificity is desired, rather
than overall prediction accuracy, the figure suggests downsampling
until the minority sample is of equal size to the majority sample.
The cost of achieving this balance is a much higher error rate in
predicting the positive class (PPV). If one wishes to balance
sensitivity with PPV, Fig. \ref{fig:downsample} suggests limited
downsampling should be performed. Downsampling reduces the
training dataset size and, thus, decreases the precision with
which model performance metrics are measured. At the minority
class sample sizes in Fig. \ref{fig:downsample}, approximately 400
students, model performance metrics are still very precisely
estimated; no data point in Fig. \ref{fig:downsample} has a
standard deviation exceeding $0.02$ or a standard error of the
mean exceeding $0.002$.

\subsubsection{Upsampling}
It is also possible to oversample or upsample the minority class
to produce a more balanced sample. This is done by randomly
replicating students in the minority class. For this sample,
upsampling was completely ineffective at producing the changes in
model performance that downsampling produced. This may be because
up-sampling does not create additional unique cases of the
minority class to train the classifier.

\subsubsection{Hyperparameter Tuning }

\begin{figure}[!htb]
    \centering
    \includegraphics[width=3.5in]{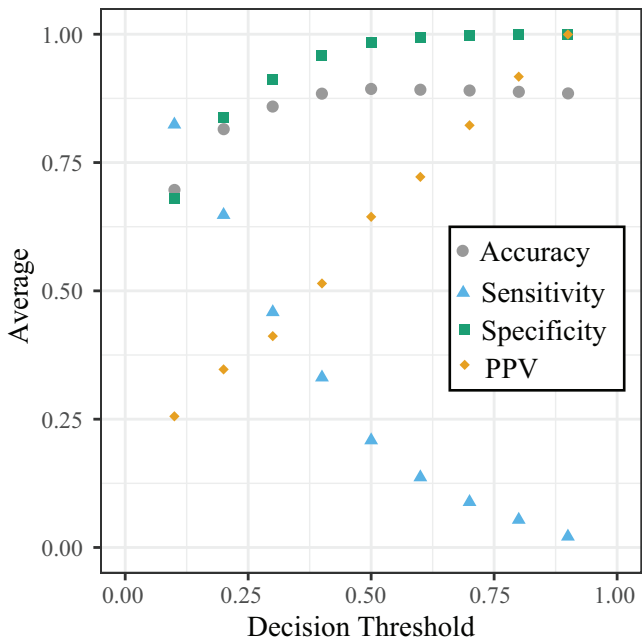}
    \caption{\label{fig:hyper} Model performance parameters plotted against the decision threshold predicting the
    DF outcome. The standard deviation did not exceed $0.01$ for any data point. }
\end{figure}

Machine learning algorithms are algorithms that are ultimately
implemented as computer programs. Like most programs, they contain
a number of parameters that can be adjusted by the user to
optimize their performance. In Study 1, the default parameters
selected by the developers of the algorithms were used. The
adjustable parameters for the random forest function in R include
the number of trees that are grown  and the decision threshold.
Random forests work by growing a large number of decision trees
and letting each tree vote on the classification. The decision
threshold sets the percentage of votes the positive classification
(CDF or DF) must receive for the individual to be classified into
that class. The default for the ``randomForest'' package in R used
in Study 1 is $50\%$. Figure \ref{fig:hyper} shows the effect of
the decision threshold on the model performance statistics
predicting the DF outcome. A decision threshold around $0.15$
provides a balance of sensitivity and specificity; however, at
this threshold the PPV is poor. A decision threshold of $0.3$
provided a balance of sensitivity and PPV (and optimizes the $F_1$
statistics). While different course applications of machine
learning may require valuing either sensitivity or PPV more
highly, for this work, we will seek to balance the two quantities
valuing both identifying the most students at risk and having this
identification be correct.

Examination of Fig. \ref{fig:hyper} allows one to understand how
sensitivity, specificity, and PPV work together. As the decision
threshold is increased, more trees have to vote for the DF outcome
for it to be selected, and therefore, fewer students are
classified as DF for higher thresholds. Because fewer students are
classified as DF, more actual DF students are misclassified by the
algorithm with higher threshold decreasing sensitivity; however,
with the more restrictive threshold, more of the predictions are
correct increasing PPV.

This analysis was repeated for the CDF outcome and a plot similar
to Fig. \ref{fig:hyper} is presented in the Supplemental Material
\cite{supp}. For the CDF outcome, a decision threshold of $0.45$
was optimal suggesting the models in Study 1 may have had a good
balance of sensitivity and PPV.

\subsubsection{Grid Search}
\label{sec:downsample} Both downsampling and tuning the decision
threshold generated models with improved classification of DF
students. The degree of downsampling can also be viewed as a
hyperparameter. In machine learning, it is not uncommon to have
multiple hyperparameters which must be optimized together to
create the best classification model. To do this, one performs a
``grid search'' through the space of hyperparameters, iterating
through combinations of hyperparameters to optimize a performance
statistic \cite{scikit}. The Supplemental Material \cite{supp}
presents contour plots of sensitivity, specificity,
$g_{\mathrm{mean}}$, $\kappa$, $F_1$, PPV, and AUC varying the
decision threshold and downsampling rate.

Figure \ref{fig:downsample} suggests limited downsampling may be
appropriate for optimizing this sample. Sensitivity, specificity,
and PPV do not have maxima on the contour plots. This was
expected; all continue to either increase or decrease with changes
in the decision threshold. AUC, $g_{\mathrm{mean}}$, and $F_1$ all
have broad maxima which include small downsampling rates. Cohen's
$\kappa$ has two narrow maxima, one of which also suggests low
downsampling rates. As such, and because downsampling eliminates
unique cases from the training data, no downsampling was performed
by the optimized classifier. It is unclear if this failure of
downsampling to improve models optimized by the decision threshold
is a general feature of grade prediction classifiers, or a unique
feature of this dataset. Researchers investigating machine
learning for student classification should explore downsampling;
however, it was not effective for the students in this sample.

Without downsampling, the decision thresholds from the previous
section (0.45 for CDF and 0.30 for DF) will be used. The 0.30
threshold is near the maximum region for all performance metrics
with a maximum.

\subsection{Unbalanced Independent Variables}
Four dichotomous variables were explored for this study: gender,
first-generation status (FirstGen), underrepresented minority
status (URM), and calculus-readiness (CalReady). Each of these
variables divide the sample into subgroups with different class
outcomes and differing levels of academic preparation as shown in
Table \ref{tab:demo}. In Study 1, demographic variables including
URM, FirstGen, and Gender were shown to have limited importance in
the prediction of whether a student would receive an A or B;
however, it was unclear to the extent that this resulted from the
highly unbalanced sample. As can be seen from Table
\ref{tab:demo}, women, first-generation students, and
underrepresented students form small subsets of the overall
sample. As with the unbalanced dependent variable, this provides
the machine learning algorithm many more examples of majority
students and possibly optimizes the prediction algorithm to the
majority class. To explore the consequences of using unbalanced
independent variables on the prediction performance of the
minority class, we first introduce an artificial independent
variable that is not co-linear with general markers of academic
preparation and success. Once this variable is understood, the
analysis is repeated for the four dichotomous independent
variables available in this dataset.

\subsubsection{An Artificial Independent Variable}
\begin{figure}[!h]
    \centering
    \includegraphics[width=3.5in]{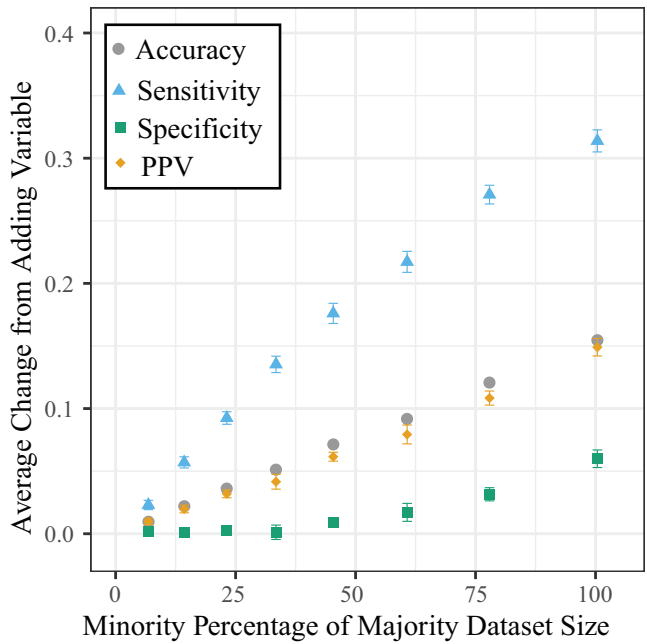}
    \caption{\label{fig:fake} Variable importance predicting CDF outcome measured by the change in model performance parameters between a
    model using an artificial dichotomous variable and one where it is removed. Error bars are one standard deviation in length. The error bars
    for the accuracy were smaller than the circle. }
\end{figure}

Table \ref{tab:demo} shows that, for the four dichotomous
variables available to this study, different levels of the
variable select students with often dramatically different
measures of high school preparation and college success. Because
the groups differ on continuous variables already included in the
analysis, it may be that the finding that the dichotomous
independent variables are not important results from these
differences. To understand the effects of variable imbalance that
is not coupled to prior academic performance and preparation, an
artificial dichotomous variable was constructed. This new variable
was randomly set to one for students with a majority outcome (AB)
and to zero otherwise. This was done by checking if a random
number was above a threshold value for majority students and
setting all minority outcome (CDF) students to zero. This variable
should be important to the prediction models because one can
perfectly predict the outcome of students when the variable equals
one. Conceptually this kind of distribution might be produced if
students were randomly assigned to a perfectly functioning
treatment so all students receiving the treatment scored an A or
B. The size of the minority class can be adjusted by changing the
threshold of the random variable. Figure \ref{fig:fake} shows the
variable importance for the artificial variable for different
minority sample sizes. The minority percentage of majority dataset
size plotted on the horizontal axis is the percentage ratio of
students for which the artificial variable is zero (the larger
class) to students where the variable is one. Variable importance
is measured by the change in some performance metric when the
variable is added to the model. This change is calculated using
bootstrapping. Figure \ref{fig:fake} clearly shows variable
importance is related to the balance of the minority and majority
classes of the artificial independent variable.

\subsubsection{Dichotomous Independent Variables}
\begin{figure*}[!htb]
    \centering
    \includegraphics[width=7in]{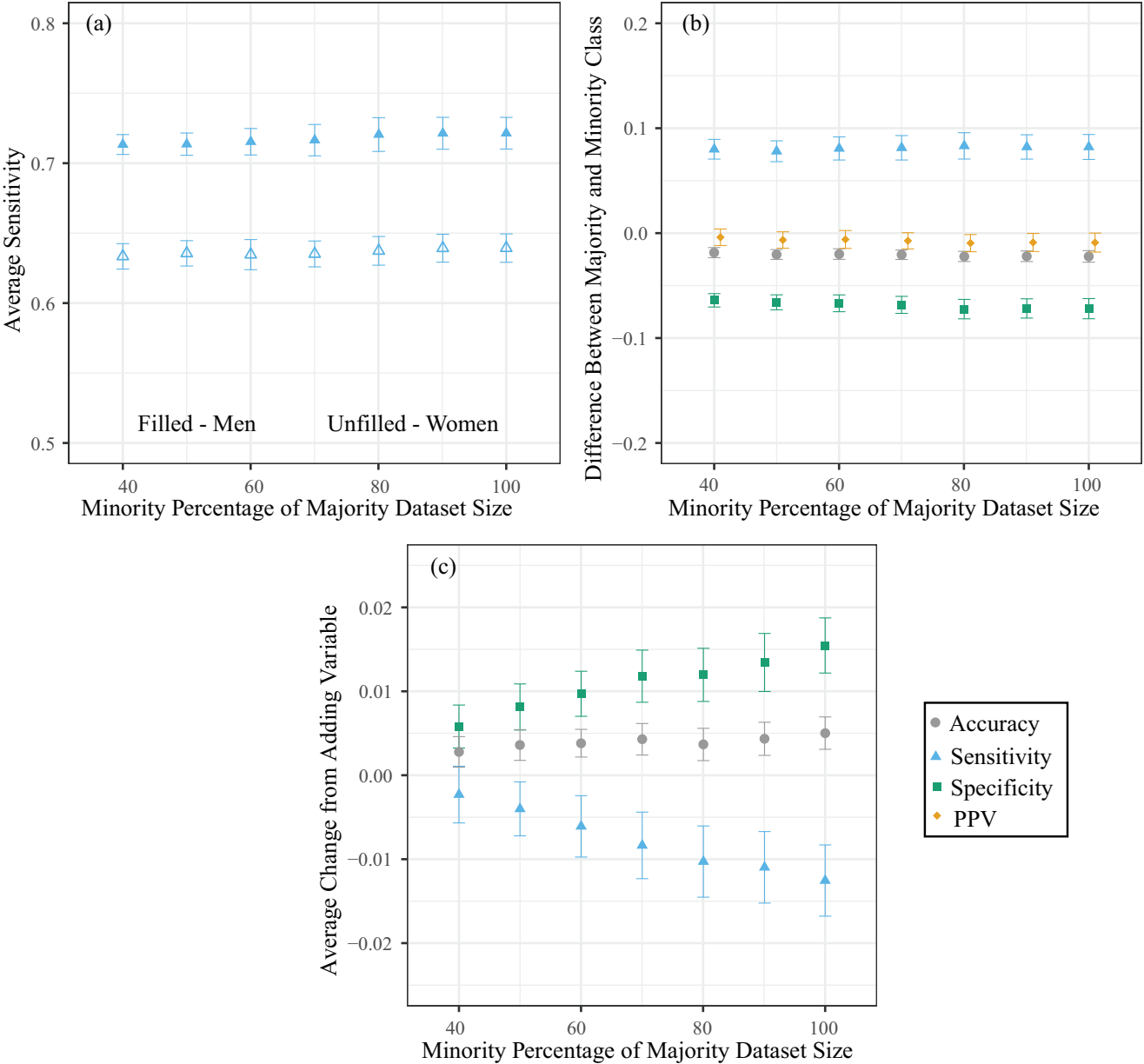}
    \caption{Model performance statistics comparing models including the gender
    variable to those which do not predicting CDF. Figure (a) shows the sensitivity of the
    model for men and women. Error bars represent one standard deviation. Figure (b) shows the difference in model performance
    between men and women. A positive difference means the model performs better for men. Figure (c) shows the change in model performance parameters as
    the gender variable is added to the model.\label{fig:gender}   }
\end{figure*}

The analysis of the previous section was repeated with each
dichotomous variable available: Gender, URM, FirstGen, and
CalReady. The results are presented in the Supplemental Material
\cite{supp}. For each variable, a plot of the sensitivity for
models including the variable and excluding the variable at
different levels of downsampling are provided. The difference in
model performance between the two classes defined by the variable
is presented, as is the difference in performance with and without
the variable. An example of these plots is shown in Fig.
\ref{fig:gender} which presents the results for the gender
variable predicting CDF outcomes. For all four variables, the
performance of the models was fairly insensitive to the level of
downsampling. As such, the conclusions about the low importance of
demographic variables in Study 1 (gender, underrepresented
minority status, and first generation status) in predicting
physics grade were not the result of the imbalance of the sample.

Figure \ref{fig:gender} shows a number of interesting features of
the effect of the inclusion of gender on model performance. The
average sensitivity (Fig. \ref{fig:gender}(a)) differs by about
10\% between men and women; the model correctly predicts when men
will receive a C, D, or F at a 10\% higher rate than women. The
accuracy and specificity of the model, however, are higher for
woman, Fig. \ref{fig:gender}(b). These difference were fairly
insensitive to the level of downsampling and, therefore, did not
result from the underrepresentation of women in the sample causing
the classifier to be predominately trained on men. Some  feature
of men or women not captured by the institutional variables in the
model must be causing the differences in classification accuracy.
Other demographic variables also demonstrated differences in some
performance metrics as shown in the Supplemental Material
\cite{supp}.

While sensitivity, specificity, accuracy, and PPV are relatively
constant at different levels of downsampling, the variable
importance of the gender variable measured as the change in these
quantities as gender is added to the model did change somewhat
with the level of downsampling, Fig. \ref{fig:gender}(c); however,
the magnitude of the change was not large and did not exceed 0.01.

\subsection{Optimal Classification Model}

\renewcommand{\tabcolsep}{2mm}
\begin{table*}[!htb]
    \caption{\label{tab:finalfit} Model performance parameters for the optimized classifier using all variables. Values represent the mean $\pm$ the standard
    deviation.}
    \begin{center}

        \begin{tabular}{|l|cccccccc|}\hline
Outcome & Accuracy          & Sensitivity       & Specificity
&PPV                & $\kappa$             & AUC               &
$F_1$          & $g_{\mathrm{mean}}$       \\\hline
\multicolumn{9}{|c|}{Overall}\\\hline
CDF     & $0.76\pm 0.01$    &  $0.69\pm 0.02$   & $0.80\pm 0.01$    & $0.67\pm 0.01$    & $0.48\pm 0.01$    & $0.74\pm 0.01$    & $0.68\pm 0.01$    & $0.74\pm 0.01$            \\
DF      & $0.87\pm 0.01$    &  $0.46\pm 0.02$   & $0.92\pm 0.01$
& $0.45\pm 0.02$    & $0.38\pm 0.02$    & $0.69\pm 0.01$    &
$0.46\pm 0.01$    & $0.65\pm 0.01$            \\\hline
\multicolumn{9}{|c|}{Female Students}\\\hline
CDF     & $0.78\pm 0.01$    &  $0.61\pm 0.03$   & $0.87\pm 0.02$    &$0.69\pm 0.03$     & $0.49\pm 0.03$    & $0.74\pm 0.02$    & $0.65\pm 0.02$    & $0.73\pm 0.02$            \\
DF      & $0.90\pm 0.01$    &  $0.39\pm 0.05$   & $0.96\pm 0.01$
&$0.49\pm 0.07$     & $0.38\pm 0.05$    & $0.67\pm 0.03$    &
$0.43\pm 0.05$    & $0.61\pm 0.04$            \\\hline
\multicolumn{9}{|c|}{Underrepresented Minority Students}\\\hline
CDF     & $0.72\pm 0.03$    &  $0.78\pm 0.04$   & $0.66\pm 0.04$    &$0.67\pm 0.04$     & $0.44\pm 0.05$    & $0.72\pm 0.03$    & $0.72\pm 0.03$    & $0.72\pm 0.03$                       \\
DF      & $0.78\pm 0.02$    &  $0.46\pm 0.08$   & $0.85\pm 0.03$
&$0.36\pm 0.06$     & $0.27\pm 0.07$    & $0.65\pm 0.04$    &
$0.40\pm 0.06$    & $0.62\pm 0.05$            \\\hline
\multicolumn{9}{|c|}{First-Generation Students}\\\hline
CDF     & $0.73\pm 0.02$    &  $0.67\pm 0.03$   & $0.77\pm 0.03$    &$0.65\pm 0.03$     & $0.44\pm 0.03$    & $0.72\pm 0.02$    & $0.66\pm 0.02$    & $0.72\pm 0.02$            \\
DF      & $0.86\pm 0.01$    &  $0.45\pm 0.06$   & $0.91\pm 0.01$
&$0.40\pm 0.06$     & $0.34\pm 0.05$    & $0.68\pm 0.03$    &
$0.42\pm 0.05$    & $0.64\pm 0.05$            \\\hline
\multicolumn{9}{|c|}{Not Calculus Ready Students}\\\hline
CDF     & $0.70\pm 0.01$    &  $0.82\pm 0.02$   & $0.55\pm 0.03$    &$0.70\pm 0.02$     & $0.38\pm 0.03$    & $0.69\pm 0.01$    & $0.76\pm 0.01$    & $0.67\pm 0.01$            \\
DF      & $0.78\pm 0.01$    &  $0.46\pm 0.04$   & $0.86\pm 0.02$
&$0.42\pm 0.03$     & $0.30\pm 0.03$    & $0.66\pm 0.02$    &
$0.44\pm 0.03$    & $0.63\pm 0.03$            \\\hline
        \end{tabular}
    \end{center}
\end{table*}

\renewcommand{\tabcolsep}{2pt}

\begin{table}[!htb]
    \caption{\label{tab:finalfitgroups} Model performance parameters fitting the subgroup only.}
    \begin{center}

        \begin{tabular}{|l|cccc|}\hline
Outcome & Accuracy          & Sensitivity       & Specificity
&PPV                       \\\hline \multicolumn{5}{|c|}{Female
Students}\\\hline
CDF     & $0.77\pm 0.01$    &  $0.66\pm 0.03$   & $0.83\pm 0.02$    &$0.65\pm 0.02$               \\
DF      & $0.89\pm 0.01$    &  $0.41\pm 0.06$   & $0.94\pm 0.02$
&$0.41\pm 0.06$                 \\\hline
\multicolumn{5}{|c|}{Underrepresented Minority Students}\\\hline
CDF     & $0.69\pm 0.02$    &  $0.73\pm 0.04$   & $0.66\pm 0.05$    &$0.65\pm 0.03$                            \\
DF      & $0.80\pm 0.02$    &  $0.35\pm 0.10$   & $0.88\pm 0.04$
&$0.37\pm 0.07$                \\\hline \multicolumn{5}{|c|}{First
Generation Students}\\\hline
CDF     & $0.71\pm 0.02$    &  $0.65\pm 0.04$   & $0.75\pm 0.03$    &$0.63\pm 0.03$                 \\
DF      & $0.86\pm 0.02$    &  $0.37\pm 0.08$   & $0.92\pm 0.02$
&$0.39\pm 0.07$                \\\hline \multicolumn{5}{|c|}{Not
Calculus Ready Students}\\\hline
CDF     & $0.72\pm 0.01$    &  $0.80\pm 0.02$   & $0.63\pm 0.02$    &$0.70\pm 0.01$                 \\
DF      & $0.79\pm 0.01$    &  $0.48\pm 0.05$   & $0.85\pm 0.02$
&$0.39\pm 0.03$                 \\\hline
        \end{tabular}
    \end{center}
\end{table}

\label{sec:optimal} With the results above, the analysis of Study
1 can be refined, and optimal classification models constructed
for the DF and the CDF outcomes. The results shown in Fig.
\ref{fig:train} show that the uncertainty in model performance
metrics decreases rapidly until the minority sample size reaches
100. We would also like to achieve accurate characterization of
the demographic subgroups in the sample; it seems likely this
threshold also applies to the subgroups. The test and training
datasets play different roles in machine learning. Machine
learning predictions generally improve as more data are used to
train the classification algorithms; therefore, as much data as
possible should be assigned to the training dataset while
retaining the minimum data required for accurate characterization
of the classifier in the test dataset \cite{scikit}. This implies
the smallest demographic subgroup controls the test-train split so
as to ideally retain at least 100 students from each subgroup in
both the test and training dataset. Table \ref{tab:demogroups}
shows there is insufficient diversity in the sample to achieve
this goal for all groups particularly for predicting the DF
outcome. Only 62 URM students and 90 first-generation students
received a D or F in the class. As such, to evenly divide these
students between the test and training dataset, a 50\% test-train
split was used. For the CDF outcome, there are only 181 URM
students in the sample who earn a C, D, or F suggesting a 50\%
test-train split. If one abandoned the goal of precisely
predicting the variable importance of this group, more data could
be retained for the training dataset.

\begin{figure*}[!htb]
    \centering
    \includegraphics[width=6in]{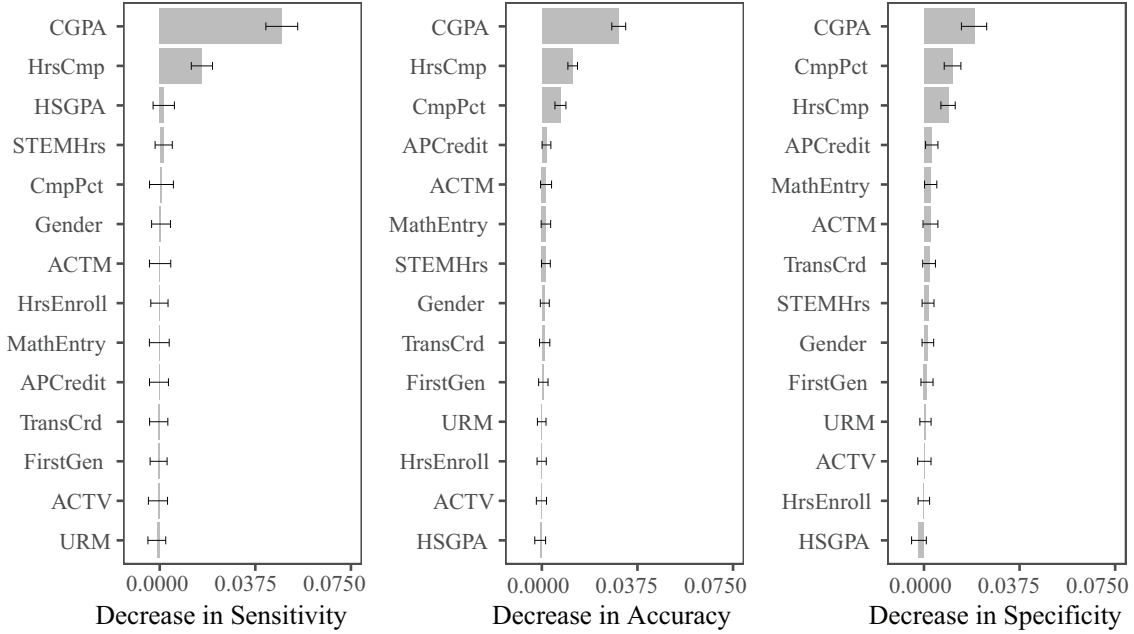}
    \caption{Variable importance of the optimized model predicting CDF. Error bars are
    one standard deviation in length. \label{fig:finalvar}   }
\end{figure*}

The analysis showing that downsampling was not productive for the
unbalanced dependent variable and the result that model
performance parameters were insensitive to the level of
downsampling for the independent dichotomous variables suggest
that downsampling is not productive for this dataset. The optimal
model is then constructed with a 50\% test-train split, no
downsampling, and the 0.30 (DF) and 0.45 (CDF) decision thresholds
suggested by hyperparameter turning. Using these parameters and
all variables in Table \ref{tab:variables} allowed the
construction of a classification model that was characterized on
all students; the model performance metrics for this model are
shown as the ``Overall'' entries in Table \ref{tab:finalfit}. This
model has a balance of sensitivity and PPV. For the CDF outcome,
the overall accuracy is slightly better than that found for only
institutional variables in Study 1 with $\kappa$ in the range of
moderate agreement, but AUC less than the threshold of 0.80 for
excellent discrimination. While the accuracy is higher for the DF
model, sensitivity and PPV are much lower. It is harder to
classify students receiving a D or F; other performance metrics
are also lower for the DF model with $\kappa$ in the range of fair
agreement.

Because no downsampling was used, the majority of the cases on
which the optimal classifier was trained do not belong to any
demographic subgroup underrepresented in physics. Because more
majority students cases were used to train the classifier, it may
perform differently for some subgroups as is suggested by Fig.
\ref{fig:gender} and figures like it for other groups in the
Supplemental Materials \cite{supp}. To investigate this
possibility, the optimal classification model was also
characterized for each minority subgroup separately as shown in
Table \ref{tab:finalfit}. This was done by using the
classification model trained on the full training set to classify
only the minority cases in the test dataset. The results are shown
in Table \ref{tab:finalfit}; differences are significant if the
standard errors of the mean do not overlap; the standard error is
the standard deviation divided by $14.1$. For the CDF outcome, the
models have substantially lower sensitivity and higher specificity
for women with approximately equal PPV consistent with Fig.
\ref{fig:gender}. Conversely, for both URM students and
non-CalReady students, the model's sensitivity is substantially
higher and specificity lower than the overall model. For
non-CalReady students, $\kappa$ is substantially lower and $F_1$
higher than for the overall model.

All DF models had substantially lower sensitivity and PPV than CDF
models; it is more difficult to predict the DF outcome for all
subgroups. The sensitivity of the DF model for women was
substantially lower than the overall model consistent with the CDF
models. For other groups, the sensitivity of the DF models were
very similar to the overall model. For URM students and
non-CalReady students, $\kappa$ was lower than the overall model;
AUC, $F_1$, and $g_{\mathrm{mean}}$ were similar.

While few substantial differences in performance metrics were
identified in Table \ref{tab:finalfit}, model performance might be
substantially different if only students from the subgroup were
used to train the model. This was investigated by training the
models on the subgroup alone; the results are shown in Table
\ref{tab:finalfitgroups}. For the CDF outcome, the sensitivity of
the model for women improved from 0.61 to 0.66 balancing a
decrease in PPV from 0.69 to 0.65. For URM students, the
sensitivity decreased for the CDF outcome. For CDF, both URM and
non-CalReady students had higher sensitivity than PPV, as they did
in the overall model; this suggests it may be appropriate to tune
the decision threshold separately for these groups. For the DF
outcome, both URM and FirstGen students had lower sensitivity when
fitted separately than when fitted in the overall model.

Overall, the optimal classification models presented in this
section achieved a balance of sensitivity and PPV as shown in
Table \ref{tab:finalfit}. Some variation existed for some
subgroups with higher sensitivity for URM students and lower
sensitivity for women. Model performance metrics were fairly
precisely estimated for all groups with a maximum standard
deviation of $0.06$ suggesting the sample size was sufficient for
effective characterization of the models; the standard error of
the mean was substantially smaller. For the CDF outcome, $\kappa$
ranges from 0.35 to 0.48, or from fair to moderate agreement; for
the DF outcome, $\kappa$ was smaller ranging from 0.23 to 0.33,
fair agreement. No AUC value met Hosmer's threshold of 0.80 for
excellent discrimination.

The importance of each variable used in the classification models
was evaluated by bootstrapping. Models were fit with the variable
and without the variable and model performance metrics compared.
The variable importance for sensitivity, specificity, and accuracy
are shown in Fig. \ref{fig:finalvar} for the overall model
predicting CDF. The variable importance for the overall model
predicting DF is similar and is presented in the Supplemental
Materials \cite{supp}. The variable importance for each
demographic subgroup is also presented in the Supplemental
Materials.

The variable importance plots shown in Fig. \ref{fig:finalvar}
show that CGPA is by far the most important variable in agreement
with Study 1. In addition to CGPA, only HrsCmp (the number of
credit hours completed) is consistently an important variable. As
in Study 1, a very limited number of institutional variables are
needed to predict grades in a physics class.

\section{Discussion}
This study sought to answer four research questions; they will be addressed
in the order proposed.

{\it RQ1: How can machine learning algorithms be applied to
predict unbalanced physics class outcomes?} Figure \ref{fig:train}
shows that the random forest algorithm using the default decision
threshold and no downsampling produces models with very low
sensitivity for a substantially unbalanced outcome variable, the
DF outcome. Model accuracy, the primary performance metric
reported in Study 1, was not an effective measure of the
performance of an unbalanced classifier. Model performance metrics
which focused on the minority outcome, the sensitivity and PPV,
were more useful in evaluating performance. Sensitivity was
substantially improved by downsampling until the minority and
majority classes were the same size; however, this somewhat
degraded the accuracy and specificity and strongly degraded PPV.
Tuning the random forest hyperparameters, specifically the
decision threshold, was also productive in increasing sensitivity,
once again at the expense of accuracy, specificity, and PPV. To
both identify as many of the minority class (the DF or CDF
outcome) measured by sensitivity and to have as large a proportion
of those identifications be accurate, measured by PPV, models that
balanced sensitivity and PPV were constructed.

A grid search allowed the identification of the combination of
downsampling and hyperparameter tuning that was optimal to produce
a balance of sensitivity and PPV. For this sample, that balance
could be achieve by adjusting the decision threshold alone without
downsampling. Downsampling, while productive in eliminating the
effects of sample imbalance, has other negative effects. By
removing cases, it lowers the number of unique individuals on
which the classifier is training, reducing performance. It also
lowers the overall training dataset size increasing the
imprecision of the performance metric estimates. With no
downsampling, the decision threshold was set to 0.30 for the DF
outcome and 0.45 for the CDF outcome; both different than 0.50
default threshold used in Study 1. Table \ref{tab:finalfit} shows
these values produced approximately equal values of sensitivity
and PPV for both the DF and CDF outcome.

At these values, the CDF outcome model was accurate in 76\% of its
predictions overall, predicting 69\% of the CDF outcomes
correctly; 67\% of students predicted to earn a CDF did so in the
test dataset. Global model fit parameters were also fairly strong
with $\kappa$ in the range of moderate agreement, and AUC near the
0.8 threshold for excellent discrimination. The DF models did not
perform nearly as well as the CDF models. While the models
predictions were accurate 87\% of the time, the models were far
more effective at predicting the ABC outcome (correctly predicted
92\% of the time) than the DF outcome (correctly predicted 46\% of
the time). The fraction of the DF predictions that were correct
was also smaller, only 45\% of the students predicted to earn a D
or F actually did. While this work did not exhaust the adjustments
that could be made to the random forest algorithm to improve DF
prediction, the results presented suggest that simply modifying
the algorithm will not be sufficient to greatly improve the
prediction accuracy of the DF model to the levels of the CDF
model. It is likely that new variables measuring different
dimensions of student motivation and performance are required to
improve prediction accuracy for students most at risk in the class
studied.

{\it RQ2: What is a productive set of performance metrics to
characterize the prediction algorithms?} This study introduced
many potential model performance metrics. Study 1 used accuracy,
AUC, and $\kappa$. This study showed that these metrics obscured a
difference in the rate of correct classification of the minority
and majority outcome classes. No single model performance
parameter was sufficient to completely understand model
performance. This work primarily used sensitivity and PPV and
sought to achieve a balance of the two quantities. This approach
focussed on the minority outcome, earning a DF or CDF, in
anticipation that identification of at-risk students may be one of
the primary applications of machine learning in PER.  Global model
fit parameters such as AUC and $g_{\mathrm{mean}}$ obscured
differences in sensitivity, specificity, and PPV and were
ineffective at distinguishing between models. $F_1$, the harmonic
mean of sensitivity and PPV, was the single parameter that most
closely aligned with the goal of optimizing both sensitivity and
PPV. Of the overall performance metrics, $\kappa$ was more
effective at distinguishing between models than AUC and
$g_{\mathrm{mean}}$; however, it has the drawback of not being
intuitively connected to the confusion matrix. As such, it is not
always clear how optimizing $\kappa$ actually influences model
predictions. The overall classification accuracy used in Study 1
was quite ineffective at distinguishing between models. This was a
result of sample imbalance, particularly for the DF outcome. One
can achieve a high accuracy by only classifying the majority class
with precision. Researchers applying classification algorithms
should not focus on a single performance measure, but should
examine a variety of measures. These measures should be chosen to
align with how the results of the classification algorithm will be
used. We suggest examining sensitivity, specificity, and PPV as a
good starting point for understanding machine learning models. One
could also compute a negative predictive value (NPV), if
predicting the successful outcome was also a focus of the model.

{\it RQ3: What sample size is required for accurate prediction of physics class outcomes?}

There was a weak increase in predictive performance with
increasing sample size until the minority sample size reached 100
as shown in Fig. \ref{fig:train}. The uncertainty in model
performance metrics decreased with increasing minority sample
size. There was not a well-defined transition, a ``knee,'' in
these plots. For the CDF outcome, the rate the uncertainty
decrease slowed at around 100 to 150 cases. For the DF outcome,
the transition to slower decline was less well defined for the
sensitivity and PPV. The minority sample size required for
commensurate uncertainty in these outcomes was also quite
different. For an uncertainty of 0.025 in the CDF outcome, 220
minority cases were needed for sensitivity and 60 cases for PPV;
for the DF outcome, 140 cases were required for sensitivity and
350 cases for PPV.

The size of the test dataset also influences the uncertainty of
the performance metrics; plots similar to Fig. \ref{fig:train} for
the test dataset are presented in the Supplemental Material
\cite{supp}. In general, larger test datasets were required to
achieve the same uncertainty as the training dataset. For an
uncertainty of 0.025 in the CDF outcome, 260 minority cases were
needed for sensitivity and 225 cases for PPV; for the DF outcome,
175 cases for sensitivity and 300 cases for PPV.

While no strong recommendation for absolute sample size can be
made, Fig. \ref{fig:train} should allow researchers wishing to
develop a classification model to determine how much uncertainty
they should expect for a given minority sample size. It should be
stressed that Fig. \ref{fig:train} plots the minority sample size
on the horizontal axis; because of differences in sample
imbalance, the overall sample size required will be quite
different for the CDF and DF outcomes for the same minority sample
size.

{\it RQ4: How does prediction accuracy differ for groups underrepresented in physics? How can machine learning
    models be optimized to predict the outcomes of all groups with equal accuracy?}

This study reported multiple model performance metrics. For the
CDF outcome, overall accuracy was somewhat, but not dramatically,
different for the demographic subgroups in the dataset varying
from 0.70 to 0.78 in Table \ref{tab:finalfit}. To produce Table
\ref{tab:finalfit}, first a model was constructed for the complete
training dataset (Overall), then that model was used to
characterize the subgroups in the test dataset. Stronger variation
was observed in other performance metrics with sensitivity varying
from 0.61 for women to 0.78 for URM students. For the DF outcome,
the variation of accuracy was similar, 0.78 to 0.90, with a
smaller variation in sensitivity; this was possibly caused by the
generally lower values of this variable. For the CDF outcome, PPV
was fairly constant for all students; more variation was observed
for DF students.

Table \ref{tab:finalfitgroups} shows that some of the differences
identified in Table \ref{tab:finalfit} decrease if models are
constructed for each demographic subgroup independently; however,
most differences remain. This indicates that the origin of the
differences in prediction performance identified in Table
\ref{tab:finalfit} were not solely the result of training the
classifier on more majority cases. It seems likely that additional
information about students in underrepresented groups may need to
be collected to produce classification models with consistent
performance across subgroups. The additional information needed is
not yet known, but may include in-class data such as homework
scores or attitudinal data such as self-efficacy or possessing a
growth mindset.

\subsection{Other Observations}
Only a few variables were important to the classification models
as shown in Fig. \ref{fig:finalvar}. There is strong theoretical
support that many of the variables identified as unimportant are
strong markers of potential academic success: HSGPA and ACTM
particularly. Table \ref{tab:demogroups} also shows strong
differences in the academic preparation of URM, FirstGen, and
non-CalReady students. It seems quite likely that all these
variables were found unimportant because their effects are already
present in the variables that were found as important: college GPA
and the number of hours completed. The variable importance would
likely change dramatically if measures of college success, CGPA,
HrsCmp, CmpPct, and STEMHrs were removed from the models leaving
only variables measured before college began.

\subsection{Recommendations for the Use of Machine Learning}
The results of this paper provide some guidance to future
researchers interested in applying machine learning algorithms to
predict physics course outcomes.
\begin{description}
\item[Test-Train Data Requirements] The results of Fig.
\ref{fig:train} suggest training datasets should be at minimum 100
students; a similar plot in the Supplemental Material suggest
similar criteria for the test dataset. However, a fixed dataset
threshold should not be used, rather bootstrapping should be
employed to establish the precision of the model performance
metrics. The decisions made with the classification model will
determine the precision needed. For models using demographic
variables for students underrepresented in physics classes,
substantially larger datasets are required to precisely
characterize model performance for these students.
\item[Performance Metrics] The application of the model
predictions should be considered before selecting performance
metrics. Some applications may value overall accuracy, while
others may be more concerned in the correct prediction of the
minority or majority outcome. This study settled on optimizing
sensitivity and PPV simultaneously; this could be accomplished by
maximizing the $F_1$ statistic. This choice focussed on the
minority outcome, receiving a DF or CDF grade, and placed equal
value on predicting as many of the minority outcomes as possible
and having the predictions be correct as often as possible.
\item[Unbalanced Outcomes] Imbalance between the majority and
minority outcome makes some performance metrics, such as accuracy
or AUC, less useful in model evaluation. An unbalanced outcome
variable can produce classifiers that are inaccurate for the
minority class, as was the DF classifier presented in this work
using the default decision threshold. Both downsampling and
hyperparameter tuning can eliminate some of the negative effects
of unbalanced outcomes. In this work, hyperparameter tuning alone
served to produce a classifier that balanced sensitivity and PPV.
\item[Unbalanced Independent Variables] Demographic subgroups
underrepresented in physics may be classified accurately less
often because the machine learning algorithm is trained on fewer
cases. The sample size of each demographic group to be included in
the model should be consider when establishing the test-train
split. Each subgroup should be examined independently to ensure
the models are performing equally for all groups. It may be
necessary to fit each group or to tune the hyperparameters for
each group separately. The differences in the performance between
groups was generally unaffected by downsampling.

\end{description}

\section{Implications}

A limited number of institutional variables are required to
construct classification models for physics classes. If physics
departments could arrange for these variables to be provided to
instructors along with tools to use these variables, at-risk
students could be identified and interventions directed toward
these students.

\section{Future}
This work focussed on the random forest machine learning
algorithm; many other algorithms exist and may provide additional
insight into student behavior and success in physics classes.

This work also considered only linear relations of the variables;
it is possible non-linear combinations of variables are important
to understand student success. This would be particularly
important if interactions between demographic variables
represented by product terms in the random forest model were
important to predicting student success.

This work focussed on institutional variables, a future study will
examine the addition of in-class variables such as homework
average and affective variables such as self-efficacy. It is
particularly important to identify the set of variables needed to
improve prediction effectiveness for the DF outcome.

This work demonstrated that the difference in model performance
for some underrepresented demographic groups could not be
explained by the imbalance of the sample alone. More research is
required to determine the additional variables needed produces
classifiers which are equally accurate for all underrepresented
students.

This worked explored many factors which might influence the
performance of machine learning classifiers using a large dataset
and a fairly small number of variables. As the number of variables
increase, additional work is needed to see how the number of
variables affects the results presented.

\section{Conclusions}

This work applied the random forest machine learning algorithm to
the prediction of student grades in an introductory,
calculus-based mechanics class. Both students receiving a D or F
and students receiving a C, D, or F were investigated. The default
parameters for the random forest algorithm produced classification
models which predicted the lower grade outcome correctly
substantially less often than the higher grade outcome. Both
downsampling and hyperparameter tuning (adjusting the decision
threshold) were productive in producing classification models
which predicted these outcomes correctly at a higher rate. When
used together, hyperparameter tuning alone produced results close
to a combination of downsampling and hyperparameter tuning without
removing data. By tuning the decision threshold, sensitivity (the
fraction of the DF or CDF outcomes classified correctly) was
improved from 20\% to 46\% for the DF outcome and from 62\% to
69\% for the CDF outcome. Three demographic subgroups were
examined in this work: women, underrepresented minority students,
and first generation students. For all subgroups, differences were
detected in model performance metrics between the subgroup
classifier and the overall classifier. These differences largely
persisted when the classification model was trained with only
members of the demographic subgroup. Some differences suggested
the classification models should be tuned independently for each
demographic group.

\begin{acknowledgments}
    This work was supported in part by the National Science Foundation
    under grant ECR-1561517 and HRD-1834569.
\end{acknowledgments}


%

\end{document}